\documentclass[twoside,twocolumn,9pt]{article}
\usepackage{extsizes}
\usepackage[super,sort&compress,comma]{natbib} 
\usepackage[version=3]{mhchem}
\usepackage[left=1.5cm, right=1.5cm, top=1.785cm, bottom=2.0cm]{geometry}
\usepackage{balance}
\usepackage{times,mathptmx}
\usepackage{sectsty}
\usepackage{graphicx} 
\usepackage{lastpage}
\usepackage[format=plain,justification=justified,singlelinecheck=false,font={stretch=1.125,small,sf},labelfont=bf,labelsep=space]{caption}
\usepackage{float}
\usepackage{fancyhdr}
\usepackage{fnpos}
\usepackage[english]{babel}
\addto{\captionsenglish}{%
  
}
\usepackage{array}
\usepackage{droidsans}
\usepackage{charter}
\usepackage[T1]{fontenc}
\usepackage[usenames,dvipsnames]{xcolor}
\usepackage{setspace}
\usepackage[compact]{titlesec}
\usepackage{hyperref}

\usepackage{epstopdf}

\definecolor{cream}{RGB}{222,217,201}

\begin{document}

\pagestyle{fancy}
\thispagestyle{plain}
\fancypagestyle{plain}{

}

\makeFNbottom
\makeatletter
\renewcommand\LARGE{\@setfontsize\LARGE{15pt}{17}}
\renewcommand\Large{\@setfontsize\Large{12pt}{14}}
\renewcommand\large{\@setfontsize\large{10pt}{12}}
\renewcommand\footnotesize{\@setfontsize\footnotesize{7pt}{10}}
\makeatother

\renewcommand{\thefootnote}{\fnsymbol{footnote}}
\renewcommand\footnoterule{\vspace*{1pt}%
\color{cream}\hrule width 3.5in height 0.4pt \color{black}\vspace*{5pt}} 
\setcounter{secnumdepth}{5}

\makeatletter 
\renewcommand\@biblabel[1]{#1}            
\renewcommand\@makefntext[1]%
{\noindent\makebox[0pt][r]{\@thefnmark\,}#1}
\makeatother 
\renewcommand{\figurename}{\small{Fig.}~}
\sectionfont{\sffamily\Large}
\subsectionfont{\normalsize}
\subsubsectionfont{\bf}
\setstretch{1.125} 
\setlength{\skip\footins}{0.8cm}
\setlength{\footnotesep}{0.25cm}
\setlength{\jot}{10pt}
\titlespacing*{\section}{0pt}{4pt}{4pt}
\titlespacing*{\subsection}{0pt}{15pt}{1pt}

\fancyfoot{}
\fancyfoot[LO,RE]{\vspace{-7.1pt}\includegraphics[height=9pt]{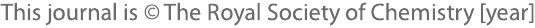}}
\fancyfoot[CO]{\vspace{-7.1pt}\hspace{13.2cm}\includegraphics{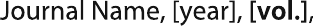}}
\fancyfoot[CE]{\vspace{-7.2pt}\hspace{-14.2cm}\includegraphics{head_foot/RF}}
\fancyfoot[RO]{\footnotesize{\sffamily{1--\pageref{LastPage} ~\textbar  \hspace{2pt}\thepage}}}
\fancyfoot[LE]{\footnotesize{\sffamily{\thepage~\textbar\hspace{3.45cm} 1--\pageref{LastPage}}}}
\fancyhead{}
\renewcommand{\headrulewidth}{0pt} 
\renewcommand{\footrulewidth}{0pt}
\setlength{\arrayrulewidth}{1pt}
\setlength{\columnsep}{6.5mm}
\setlength\bibsep{1pt}

\makeatletter 
\newlength{\figrulesep} 
\setlength{\figrulesep}{0.5\textfloatsep} 

\newcommand{\topfigrule}{\vspace*{-1pt}%
\noindent{\color{cream}\rule[-\figrulesep]{\columnwidth}{1.5pt}} }

\newcommand{\botfigrule}{\vspace*{-2pt}%
\noindent{\color{cream}\rule[\figrulesep]{\columnwidth}{1.5pt}} }

\newcommand{\dblfigrule}{\vspace*{-1pt}%
\noindent{\color{cream}\rule[-\figrulesep]{\textwidth}{1.5pt}} }

\makeatother

\twocolumn[
  \begin{@twocolumnfalse}
\vspace{3cm}
\sffamily
\begin{tabular}{m{4.5cm} p{13.5cm} }

\includegraphics{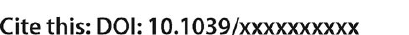} & \noindent\LARGE{\textbf{Spectroscopy of corannulene cations in helium nanodroplets$^\dag$}} \\
\vspace{0.3cm} & \vspace{0.3cm} \\

 & \noindent\large{Michael Gatchell,$^{\ast}$\textit{$^{a,b}$} Paul Martini,\textit{$^{a}$}  Felix Laimer,\textit{$^{a}$}   Marcelo Goulart,\textit{$^{a}$} Florent Calvo,\textit{$^{c}$} and Paul Scheier\textit{$^{a}$}} \\

\includegraphics{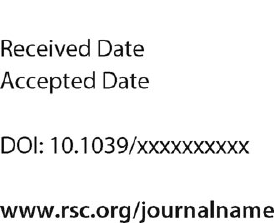} & \noindent\normalsize{Helium tagging in action spectroscopy is an efficient method for measuring the absorption spectrum of complex molecular ions with minimal perturbations to the gas phase spectrum. We have used superfluid helium nanodroplets doped with corannulene to prepare cations of these molecules complexed with different numbers of He atoms. In total we identify 13 different absorption bands from corannulene cations between 5500\,\AA\ and 6000\,\AA. The He atoms cause a small, chemically induced redshift to the band positions of the corannulene ion. By studying this effect as a function of the number of solvating atoms we are able to identify the formation of solvation structures that are not visible in the mass spectrum. The solvation features detected with the action spectroscopy agree very well with the results of atomistic modeling based on path-integral molecular dynamics simulations. By additionally doping our He droplets with D$_2$, we produce protonated corannulene ions. The absorption spectrum of these ions differs significantly from the case of the radical cations as the numerous narrow bands are replaced by a broad absorption feature that spans nearly 2000\,\AA\ in width.
} \\

\end{tabular}

 \end{@twocolumnfalse} \vspace{0.6cm}

  ]

\renewcommand*\rmdefault{bch}\normalfont\upshape
\rmfamily
\section*{}
\vspace{-1cm}


\footnotetext{\textit{$^{a}$~Institut f\"{u}r Ionenphysik und Angewandte Physik, Universit\"{a}t Innsbruck, Technikerstr.~25, A-6020 Innsbruck, Austria. E-mail: michael.gatchell@uibk.ac.at}}
\footnotetext{\textit{$^{b}$~Department of Physics, Stockholm University, 106 91 Stockholm, Sweden.}}
\footnotetext{\textit{$^{c}$~Universit\'{e} Grenoble Alpes, CNRS, LIPhy, Grenoble, France. }}

\footnotetext{\dag~Electronic Supplementary Information (ESI) available: Atomic coordinates and partial charges of corannulene cations used in our simulations and solvation dependent behavior of additional absorption bands. See DOI: 10.1039/b000000x/}



\section*{Introduction} \label{sec:intro}
Messenerger spectroscopy\cite{Okumura:1985aa} with helium atoms as tags has in recent years paved the way for determining the previously unobtainable gas phase absorption spectra of complex molecules with the accuracies required for identifying the origin of astronomical absorption features. This was demonstrated particularly well in 2015 when C$_{60}^+$ was shown \cite{Campbell:2015aa,Campbell:2016ab,Walker:2015aa} to be the first identified carrier of at least five of the several hundred Diffuse Interstellar Bands (DIBs), the origin of which have been a mystery for nearly a century.\cite{Hobbs:2009aa} Maier and coworkers achieved this result by storing the C$_{60}^+$ ions in an ion trap cooled to 4\,K. The trapped ions interacted with He gas so that complexes of He$_n$C$_{60}^+$ ($1\leq n\leq3$) were formed. These complexes were irradiated with a tunable laser and when a photon was absorbed by the fullerene ion, the excess energy would cause any attached He to boil off. By measuring the depletion of He$_n$C$_{60}^+$ using a mass spectrometer as a function of wavelength and cluster size, the gas phase spectrum of bare fullerene ions could be deduced through interpolation down to $n = 0$.\cite{Hardy:2015aa,Campbell:2015aa,Jasik:2013aa,Campbell:2016aa} The key benefit of using helium for spectroscopy in this way is the extremely weak interaction between He atoms and the molecule being studied that gives a relatively small shift compared to other methods such as matrix isolation in solid neon and argon.\cite{Fulara:1993aa,Foing:1994aa}

One of the difficulties in performing the experiments that led to the identification of C$_{60}^+$ as a DIBs carrier was attaching the He atoms to the fullerene ions. Helium is the most chemically inert element and the binding energy of a He atom to C$_{60}^+$ is only about 10\,meV.\citep{PhysRevLett.108.076101} However, droplets of He, containing thousands or millions of atoms, can efficiently capture and swallow gas phase dopants, cooling them to the 0.37\,K equilibrium temperature of the surrounding He.\cite{Toennies:2004aa,Mauracher:2018aa} By impacting these doped droplets with energetic electrons, both positively and negatively charged products can be formed.\cite{Mauracher:2018aa} This leads to charged dopants being expelled from the droplets, often complexed with some He atoms still attached, so that they can be studied using mass spectrometry.\cite{Toennies:2004aa,Mauracher:2018aa} Compared to measuring the spectra of molecular systems fully solvated in He droplets,\cite{Stienkemeier:2006aa} this method of producing ions tagged with a relatively small number of He atoms allows for the distortions to the electronic transitions to be minimized and the effects to be quantified as a function of the level of solvation.

\citet{Kuhn:2016aa} used helium nanodroplets to produce C$_{60}^+$ decorated with up to about 100 He atoms on which spectroscopy measurements were performed. These showed that the absorption lines of He$_n$C$_{60}^+$ were systematically redshifted by 0.72(1)\,\AA\ per He atom for $5 \leq n\leq 32$,\cite{Kuhn:2016aa} in good agreement with the findings of \citet{Campbell:2016ab} For $32 <  n \leq 60$ a slight blueshift per He atom was observed before the lines again were redshifted towards the positions for C$_{60}^+$ solvated in bulk He. These observations were explained by the first 32 He atoms occupying essentially localized positions above the 20 hexagonal and 12 pentagonal faces of the fullerene cages, each providing a nearly equally strong influence to the electronic structure of the C$_{60}^+$.\cite{Kuhn:2016aa,PhysRevLett.108.076101} After that the subsequent He atoms continue to fill up the first monolayer of He adsorbed to the fullerene, displacing other atoms from their preferred positions above the pentagonal rings until this layer was filled at $n=60$.\cite{Kuhn:2016aa,PhysRevLett.108.076101} This leads to a general weakening of the interaction between the individual He atoms and the fullerene ion, giving a slight blueshift with each additional He. For $n> 60$ a second solvation shell of He begins to form which has a much weaker interaction with the C$_{60}^+$ cage.\cite{Kuhn:2016aa,PhysRevLett.108.076101} This picture, together with the statistical sampling of different isomers of the He$_n$C$_{60}^+$ complexes from path-integral molecular dynamics (PIMD) simulations was also used to explain observed non-monotonic variations in the width of the 9635\,\AA\ line as a function of $n$.\cite{Kaiser:2018aa}

Recently, \citet{Hardy:2017fx} presented measurements of the absorption spectra of coronene (C$_{24}$H$_{12}$) and corannulene (C$_{20}$H$_{10}$) cations, two species of polycyclic aromatic hydrocarbons (PAHs) considered as potential DIBs carriers, using the He tag technique. While that study produced a negative result---the measured absorption bands do not match any known DIBs---they observed that C$_{20}$H$_{10}^+$ ions decorated with one and two He atoms had resonance energies at the same position, i.e.\ no blue- or redshift was observed, and concluded that red-shift induced by the He tagging to be approximately 0.1\,\AA,\cite{Hardy:2017fx} significantly smaller than the shift observed for C$_{60}^+$.

Here we present measurements of the absorption spectrum of corannulene cations, produced from doped He nanodroplets, using the He messenger technique for systems containing up to 60 He atoms. In this way we have identified 13 absorption bands in the range of 5500 to 6000\,\AA. Compared to the spectrum of C$_{60}^+$ decorated with He, where lines are redshifted linearly as a function of the number of solvation atoms, we find a much more complex behavior with corannulene with a varying magnitude of chemical shift being induced by each additional He atom. These features are well explained by atomistic modeling using path-integral molecular dynamics simulations which accurately sample the delocalized nature of the weakly interacting He atoms. Spectroscopic measurements of He-tagged molecular ions proves to be a highly sensitive method for probing the structures of weakly bound complexes and it allows us to identify changes in the solvation structures that are not accessible by conventional mass spectrometry alone.

\section*{Experimental Setup} \label{sec:exp}
Neutral helium nanodroplets are formed in the supersonic expansion of compressed He gas (2.5 MPa, 99.9999\% purity) through a 5\,$\mu$m nozzle that is cooled to 9.5\,K. This results in superfluid He droplets with an equilibrium temperature of 0.37\,K that have a typical size of a few million atoms.\cite{Knuth:1999aa} The central part of the beam passes through a 0.8\,mm skimmer located 8\,mm downstream. The droplets then transverse a series of pickup chambers where the droplets are doped with corannulene (evaporated from an oven at 70$\,^\circ$C) and D$_2$ gas (99.8\% purity, fed from a bottle through a gas line in the pickup chamber in experiments where D$_2$ is used). The doped droplets are ionized by the impact of 46\,eV electrons which mainly forms He$^+$ near the surface of the droplet. The charge migrates through the droplet by resonant hole hopping,\cite{Buchenau:1991aa} attracted by the dopant that is more polarizable than the surrounding He. The large difference in ionization energy between He and the dopant leads to the latter being ionized and expelled from the droplet, often together with some additional neutral He atoms. This leaves a distribution of dopant ions with up to several tens of He atoms (or D$_2$ molecules) attached upon which spectroscopy is performed. The ionized systems are then deflected 90$^\circ$ towards the entrance of the time-of-flight mass spectrometer (Tofwerk AG model HTOF) that is mounted perpendicularly to the neutral beam line. The deflection of the charged beam ensures that the laser does not overlap with the neutral molecular beam. 

The mass spectrometer operates on a 10\,kHz cycle and before every tenth extraction pulse, a tunable OPO (Optical Parametric Oscillator, EKSLPA model NT242 with up to 450\,{$\mu$}J per pulse) laser with a 1\,kHz repetition rate is used to probe the decorated corannulene ions in the region between the ion source and the mass spectrometer. During the remaining nine out of ten cycles of the mass spectrometer, a background spectrum is accumulated with the laser switched off. This background spectrum is then used to flatten the measured absorption spectrum to correct for the small variations in the current of ions over time.

The wavelength of the laser is scanned in steps of 0.1\,\AA\ (5\,cm$^{-1}$ linewidth) over the measurement window and mass spectra are recorded with a $m/\Delta m$ resolution of approximately 2000 for each step. The absorption spectra of He$_n$C$_{20}$H$_{10}^+$ (or D$_n$C$_{20}$H$_{10}^+$) complexes are determined by measuring the depletion of ions of a given mass as a function of wavelength as they boil off He atoms (D$_2$ molecules) following the absorption of photons.

\section*{Simulation Details} \label{sec:theory}

Classical and quantum structures of He$_n$C$_{20}$H$_{10}^+$ clusters
were theoretically determined following an already established
computational framework.\cite{Calvo:2015aa} Under the experimental
condition, the corannulene cation is essentially frozen in its
rovibrational ground state and its geometry was obtained from a
standard electronic structure calculation employing the B3LYP
functional and the 6-311++G** basis set. In our modeling, helium atoms
interact with the corannulene cation through pairwise
repulsion-dispersion forces of the Buckingham type, and a vectorial
polarization contribution is included as well. Here the electric field
on individual helium atoms results from the partial charges
distributed on the molecular cation, which were obtained using a RESP (Restrained ElectroStatic Potential)
procedure at the optimized geometry. The geometry and partial charges
employed in the calculations are provided as electronic supplementary
information.

The stable structures of He$_n$C$_{20}$H$_{10}^+$ clusters were first
explored using basin-hopping global optimization, which relies on a
Monte Carlo exploration of locally minimized geometries.\cite{Wales:1997aa}
Here, in addition to conventional (large amplitude) Monte Carlo moves,
we also implemented a specific mirror displacement move in which a
specific atom was moved to the other side of the corannulene with a
fixed probability (5\%). Such random displacements are especially
useful at small sizes, where successfully moving across the large
hydrocarbon is unlikely. Ten series of $10^5$ collective MC moves were
thus performed and generated putative classical global minima for $n$
ranging from 1 to 60.

Nuclear delocalization was then accounted for by performing
PIMD simulations, employing as in
our earlier investigation \cite{Calvo:2015aa} a Trotter discretization
number of $P=64$ at temperature $T=1$\,K, which is high enough for
exchange effects to be safely ignored. The PIMD equations of motion
were numerically solved using a velocity Verlet algorithm with a time
step of 1\,fs and the temperature being imposed using a massive Nos\'e-Hoover
thermostat with mass $3n/k_{\rm B}T$ which is the usually recommended
value.\cite{Perez:2009aa} The PIMD trajectories were integrated for
1.5\,ns, with statistical averages accumulated after the first 500~ps
equilibration period. From these simulations, the quantum energies
were evaluated from the standard virial definition,\cite{Perez:2009aa} and
for ease of visualization we also recorded the three-dimensional
atomic densities.

\section*{Results and Discussion} \label{sec:res}
\subsection*{Absorption spectrum of corannulene complexed with He}

\begin{figure*}[t] 
   \centering
   \includegraphics[width=7in]{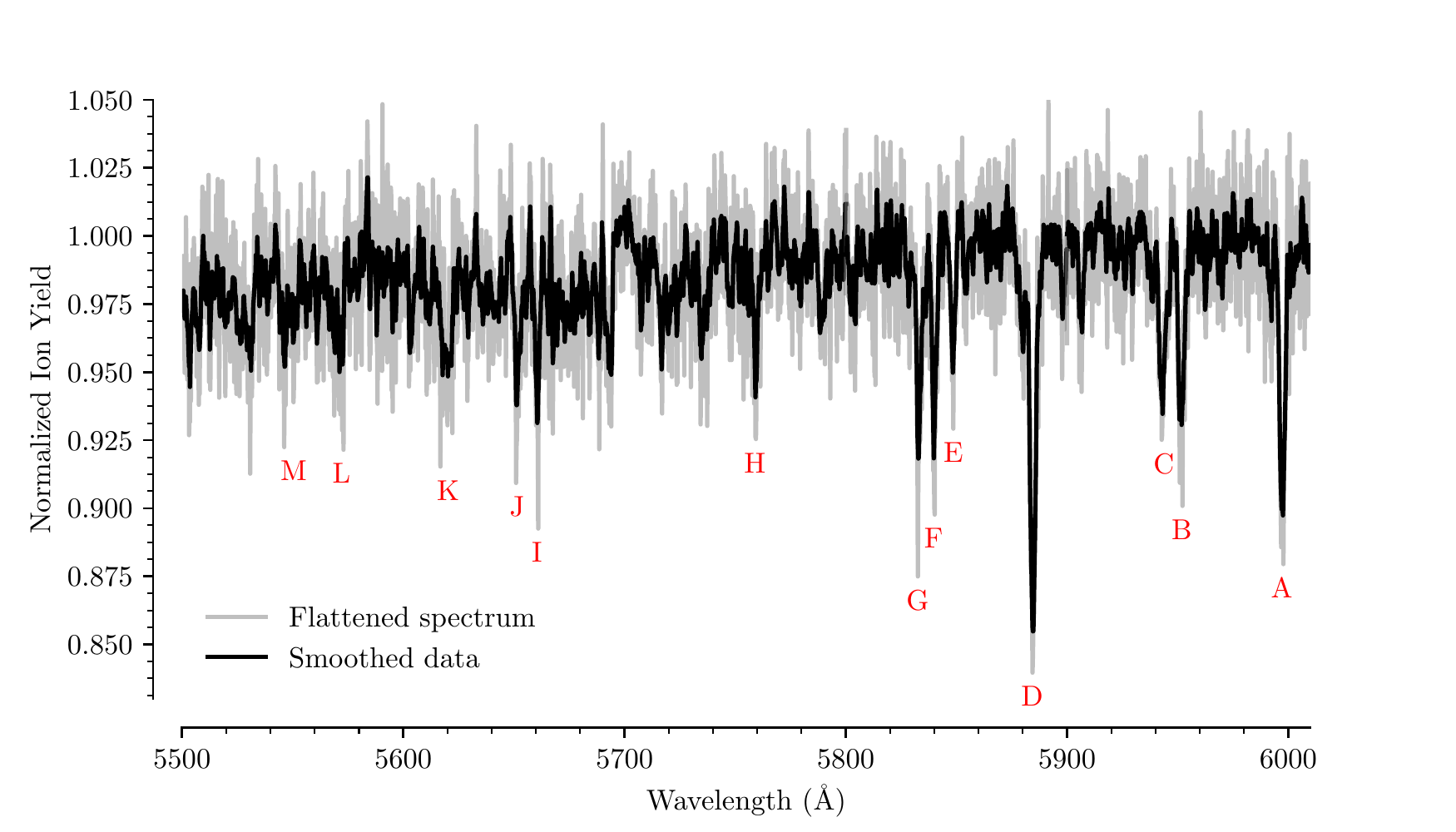} 
   \caption{Absorption spectrum in the range from 5500\,\AA\ to 6000\,\AA\ of C$_{20}$H$_{10}^+$ complexed with a single He atom. The flattened raw spectrum is shown in gray and is overlaid with a spectrum where the data has been smoothed (moving average over 10 nearest neighbors) for easier viewing. All of the analysis is performed on the unsmoothed dataset. Identified absorption lines are labeled and their positions are reported in Table \ref{tab:lines}.} 
   \label{fig:OV}
\end{figure*}

Figure \ref{fig:OV} shows an overview of the HeC$_{20}$H$_{10}^+$ absorption spectrum detected by the depletion of ions of mass 254\,u as a function of laser wavelength in the range of 5550 to 6000\,\AA. In this range we have identified 13 absorption features that we can reliably fit with Gaussian functions for several different numbers of attached He atoms. These are labeled from A through M in Figure \ref{fig:OV}. Other minima that appear to be visible in Figure \ref{fig:OV}, but which are not found in the spectra of systems containing more He atoms, are not considered for further analysis. Bands A, B, C, D, F, and G were reported by \citet{Hardy:2017fx} as two series of bands originating from the Jahn-Teller distorted electronic ground state of C$_{20}$H$_{10}^+$, which they assigned to S$_3$($^2$A$^\prime$) $\leftarrow$ S$_0$($^2$A$^{\prime\prime}$) and S$_3$($^2$A$^{\prime\prime}$) $\leftarrow$ S$_0$($^2$A$^\prime$) transitions, respectively. We do not conclusively assign the other seven bands that we identify, but expect that vibrational excitations of the S$_3$ state contributes to them. Calculations show that transitions to the S$_4$($^2$A$^\prime$)/S$_4$($^2$A$^{\prime\prime}$) states may also be responsible for some of these bands.\cite{Rice:2015aa}

\begin{figure}[t] 
   \centering
   \includegraphics[width=3.5in]{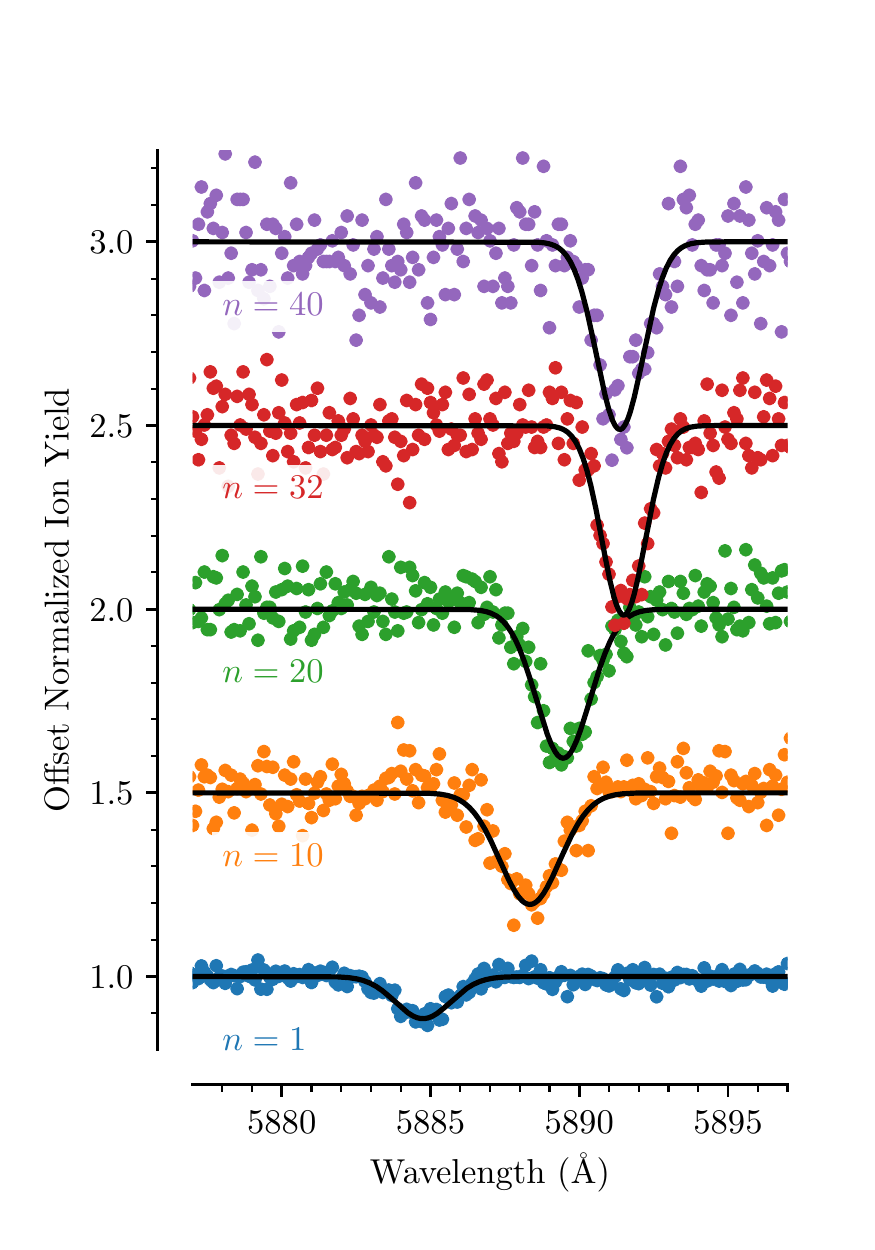} 
   \caption{The absorption feature of He$_n$C$_{20}$H$_{10}^+$ near 5885\,\AA\ (line D) shown for different numbers of $n$. The line is redshifted in a non-linear fashion up to $n=32$, after which the position is very slightly blue-shifted again. Gaussian fits are shown as solid black lines.} 
   \label{fig:fits}
\end{figure}

For each of the bands labeled in Figure \ref{fig:OV} we have performed a least square fit of a Gaussian line profile. This was then repeated for the corresponding spectra with more He atoms attached to the C$_{20}$H$_{10}^+$ ions. Figure \ref{fig:fits} shows a few examples of such fits to band D near 5885\,\AA\ and the positions and relative intensities of all bands (for the HeC$_{20}$H$_{10}^+$ complex) are given in Table \ref{tab:lines}. From Figure \ref{fig:fits} it is clear that the band position is systematically redshifted by the addition of solvating He atoms. This trend continues up to $n=32$ before the positions are slightly blueshifted again.

\begin{table}[t]
\small
\caption{Observed absorption bands for He$_n$C$_{20}$H$_{10}^+$ complexes with measured values reported for $n=1$. The given uncertainties are the statistical errors from the least square fits of Gaussian profiles to the absorption features. \label{tab:lines}}
  \begin{tabular*}{0.48\textwidth}{@{\extracolsep{\fill}}ccc}
\hline
Band & Center Wavelength (\AA) & Relative Strength \\
\hline
A & $5997.06 \pm 0.08$ & $0.86 \pm 0.06$ \\
B & $5951.86 \pm 0.18$ & $0.34 \pm 0.06$ \\
C & $5943.75 \pm 0.17$ & $0.28 \pm 0.06$ \\
D & $5884.21 \pm 0.08$ & $1.00 \pm 0.07$ \\
E & $5848.60 \pm 0.44$ & $0.32 \pm 0.10$ \\
F & $5839.68 \pm 0.21$ & $0.27 \pm 0.07$ \\
G & $5832.51 \pm 0.16$ & $0.42 \pm 0.07$ \\
H & $5758.97 \pm 0.08$ & $0.76 \pm 0.04$ \\
I & $5660.43 \pm 0.14$ & $0.27 \pm 0.05$ \\
J & $5651.43 \pm 0.14$ & $0.36 \pm 0.05$ \\
K & $5620.23 \pm 0.17$ & $0.48 \pm 0.06$ \\
L & $5572.13 \pm 0.34$ & $0.18 \pm 0.05$ \\
M & $5550.43 \pm 0.83$ & $0.19 \pm 0.07$ \\
\hline
  \end{tabular*}

\end{table}

\begin{figure*}[] 
   \centering
   \includegraphics[width=7in]{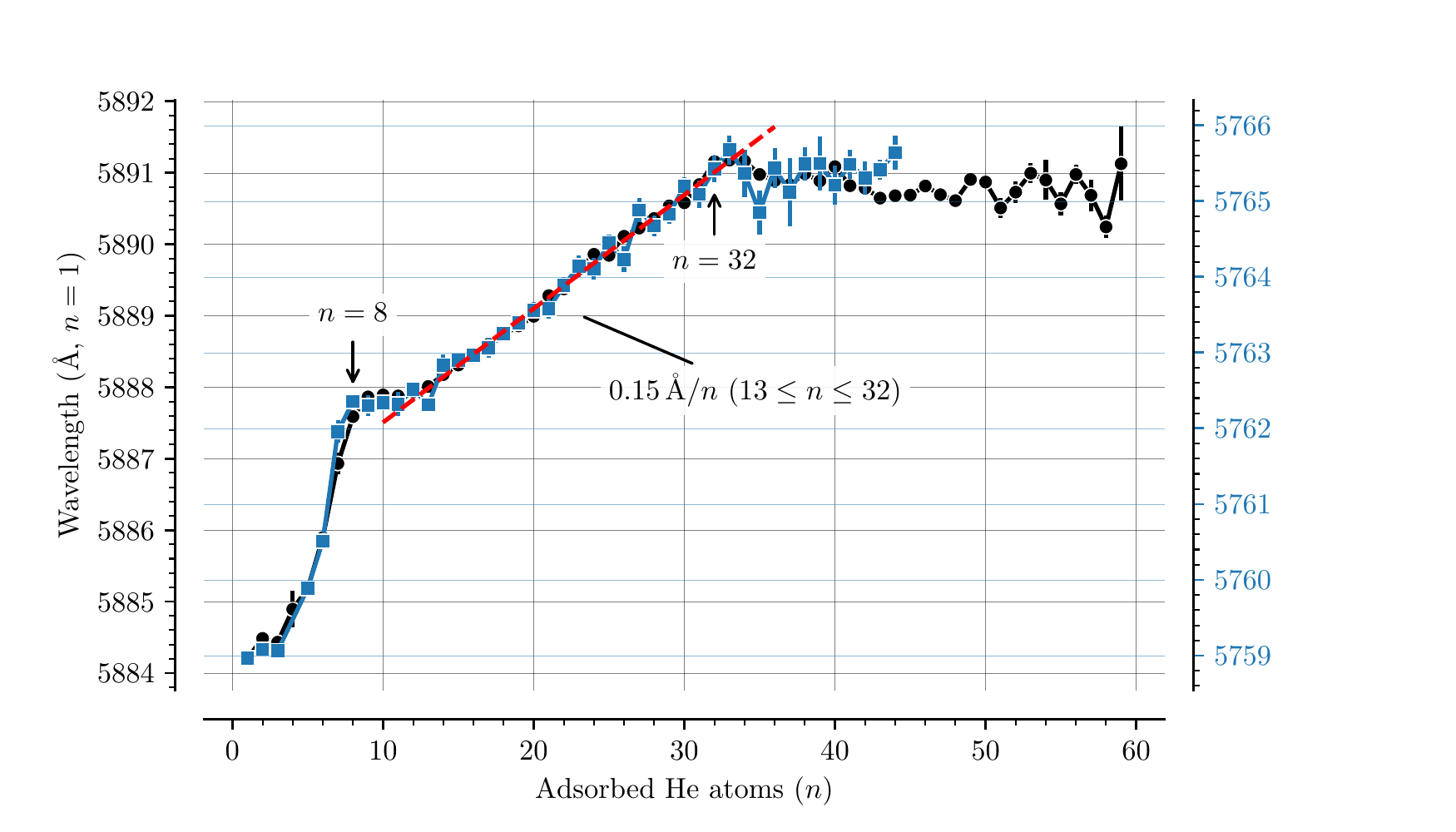} 
   \caption{Absorption wavelengths of bands D (black circles, left scale) and H (blue squares, right scale) as a function of the number of He atoms ($n$) adsorbed on the corannulene cations. Both bands show the same behavior. A linear redshift trend is observed in the $13\leq n \leq 32$ range together with a fit (red dashed line). The $n=4$ data point in the blue data set is missing due to the large uncertainty caused by corannulene complexed with water which has a similar mass. All of the remaining bands that are studied show the same behavior and are presented in the electronic supplementary information.} 
   \label{fig:trend}
\end{figure*}

Figure \ref{fig:trend} shows how the central positions of two absorption bands (D and H) vary as a function of the number of complexing He atoms and the data for the remaining bands are shown in the electronic supplementary information. For the first three He atoms, there are only small changes in the resonance wavelengths of the He$_n$C$_{20}$H$_{10}^+$ complex, with less than 0.2\,\AA\ separating them. After this there is strong redshifting in each step between $n=3$ and $n=8$, where the absorption bands are shifted by, on average, 0.7\,\AA\ per He atom (however the trend is not entirely linear). A plateau is then observed in the range between $n=8$ and $n=13$ where the wavelengths remain essentially constant. This is then followed by the $13 \leq n \leq 32$ region where the wavelength is redshifted following a stable linear trend of 0.15\,\AA\ per He atom. For $n>32$ the wavelengths remain nearly constant with a slight blueshift relative to the $n=32$ band positions. All 13 of the studied absorption features show the same basic behavior and no systematic variations in the band widths are observed as a function of the number of He atoms for any line.

The trend observed with corannulene-He complexes differs significantly compared to that observed with C$_{60}^+$-He complexes. In the latter a constant redshift of 0.7\,\AA\ per He atom was observed for the first 32 He atoms.\cite{Kuhn:2016aa,Campbell:2016ab} This persistent trend was explained by the spherical symmetry of the fullerene cages and its 32 faces, above which He atoms are preferably located,\cite{PhysRevLett.108.076101} and by utilizing a model to describe the polarization of the fullerene cage induced by the surrounding He atoms.\cite{Kuhn:2016aa} Because of the high symmetry, all of the first 32 He atoms had nearly the same interaction energy with the C$_{60}^+$, thus inducing the same spectral shift. The linear trend also meant that the gas phase spectrum of the bare C$_{60}^+$ ion could be estimated by a linear extrapolation down to zero He atoms.\cite{Campbell:2015aa} For corannulene, determining the unperturbed gas phase spectrum with the same accuracy as for C$_{60}^+$ presents a difficulty because of the more complex nature in how the He atoms influence the band positions.

\subsection*{Structures of corannulene complexed with He}

\begin{figure*}[t] 
   \centering
   \includegraphics[width=6in]{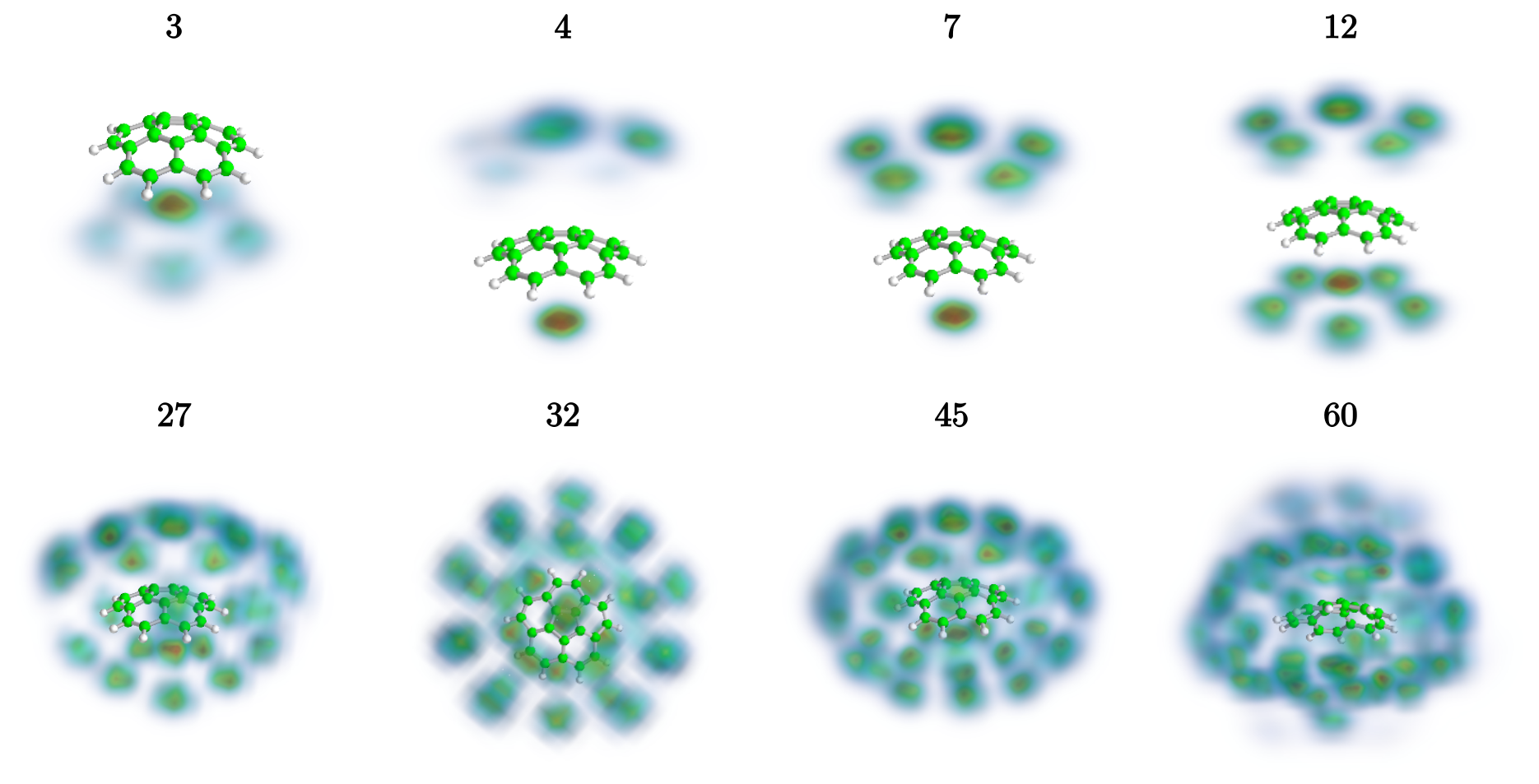}   
   \caption{Probability densities for the most stable geometries of He-solvated corannulene cations for different numbers of He atoms from our atomistic modeling. The first and most tightly bound He atom is preferentially positioned over the pentagonal face on the concave side of the corannulene ion.} 
   \label{fig:PIMDstrucs}
\end{figure*}
\begin{figure*}[t] 
   \centering
   \includegraphics[width=6in]{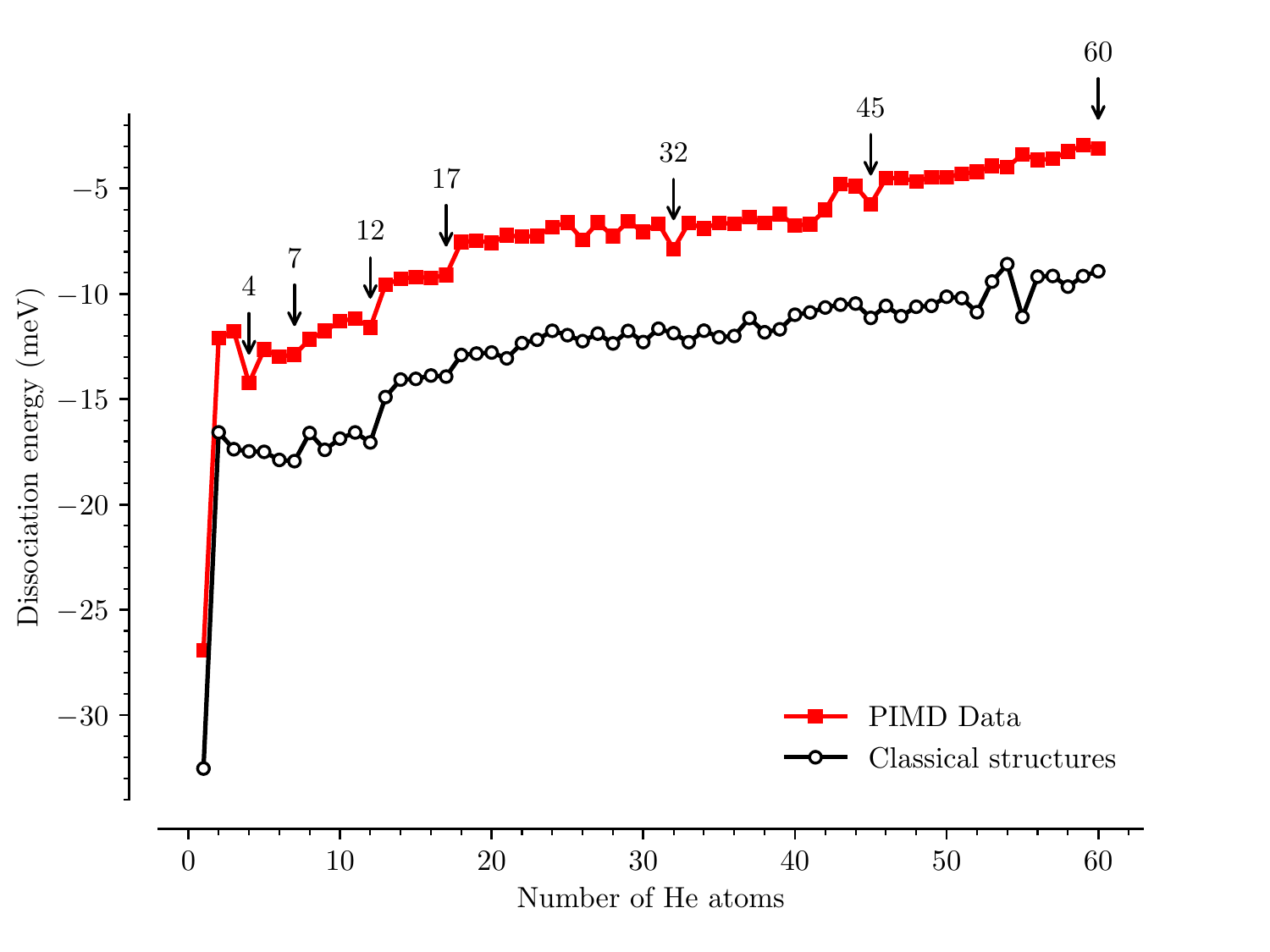}   
   \caption{Individual binding energies of He atoms physisorbed to the corannulene cation as a function of of the number of atoms from our path-integral molecular dynamics simulations (red squares) and the equivalent classical structures (black circles). The PIMD simulations consider the delocalized quantum nature of the He atoms on the shallow potential energy surface and provide a more accurate description of the interactions than the classical calculations.} 
   \label{fig:PIMDens}
\end{figure*}

We have used path-integral molecular dynamics simulations in order to investigate the solvation of the corannulene cation in He and explain the features seen in Figure \ref{fig:trend}. Density plots for the locations of He atoms surrounding a corannulene cation for different degrees of solvation are shown in Figure \ref{fig:PIMDstrucs} and the corresponding binding energies are shown in Figure \ref{fig:PIMDens}. The simulations show that the first He atom that binds to C$_{20}$H$_{10}^+$ preferentially occupies the position above the central carbon pentagon of the concave side of the molecule. The binding energy at this site is found to be about 27\,meV, which is considerably higher than for positions above the pentagonal and hexagonal faces on the convex side of the molecule that are equivalent to the outer surface of a C$_{60}^+$ ion. The next two additional He atoms also preferentially occupy the inside surface of the corannulene bowl, but due to the curvature and the interaction between the He atoms, they are not strongly localized and can easily move around the center He atom along the inner rim. This competition for space in the corannulene ion leads to the mean interaction energy per He sharply decreasing for the first three He adducts as seen in Figure \ref{fig:PIMDens}.

For complexes with between four and seven He atoms, the most favorable structures all involve a single He atom near the center of the concave side, with the remaining He atoms now occupying the convex side. Like with He$_n$C$_{60}^+$ complexes, the He atoms on the convex outer surface of the corannulene cation are each preferentially located above the faces of the carbon rings. Once the six outer faces are occupied, the filling then continues on the concave side. From $n=8$ to $n=12$, the rim of the concave side is filled with the He atoms mainly occupying positions above the open HCCCH pockets. For $n \geq 13$, the additional He atoms fill the periphery of the corannulene ion, again starting with the convex side, until the first, essentially solid, solvation layer is entirely filled at $n=32$. After this, additional He atoms may slot into the first solvation layer by displacing some of the localized He from their preferred positions. We do not identify a sharp closure of this sub-layer in the simulations, but it occurs close to $n=45$. For sizes larger than this, we find that a second solvation layer begins to form, with the second shell's He atoms largely being shielded from the corannulene by the first layer of He. In the last structure of Figure \ref{fig:PIMDstrucs} ($n=60$) we can see that a second, more delocalized layer of He has been formed with the outer layer of He mainly occupying the shallow dimples between He atoms in the first layer.

The changes in solvation structures determined by the simulations (Figures \ref{fig:PIMDstrucs} and \ref{fig:PIMDens}) describe very accurately the features seen in the solvation-dependent band positions (Figure \ref{fig:trend}). The small differences in band positions for He$_n$C$_{20}$H$_{10}^+$ complexes with $1 \leq n \leq 3$ are consistent with the first three He atoms occupying the concave side of the corannulene ion and competing for space, so that the total interaction energy in the complex remains nearly constant as the second and third He atoms are added. Similarly, following the complete covering of the carbon faces on the convex side at $n=7$, the plateau seen in Figure \ref{fig:trend} for $8 \leq n \leq 13$ is again well explained by the filling of the concave side of the corannulene ion in the simulations. Finally, the closure of the first solvation shell at $n=32$ is clearly seen in both the spectroscopic data and the simulations. This value is consistent with results for the slightly larger coronene cation, which consists of seven hexagons in a planar structure, where the equivalent solvation shell forms after 38 He atoms.\cite{Rodriguez-Cantano:2015aa,Kurzthaler:2016aa} However, the fact that He$_n$C$_{60}^+$ also shows a form of shell closure (with an associated maximum in the redshift of absorption bands) at the very same size of $n=32$ is merely a coincidence.\cite{Kuhn:2016aa,PhysRevLett.108.076101}

\subsection*{Comparison with results from mass spectrometry}

\begin{figure*}[] 
   \centering
   \includegraphics[width=7in]{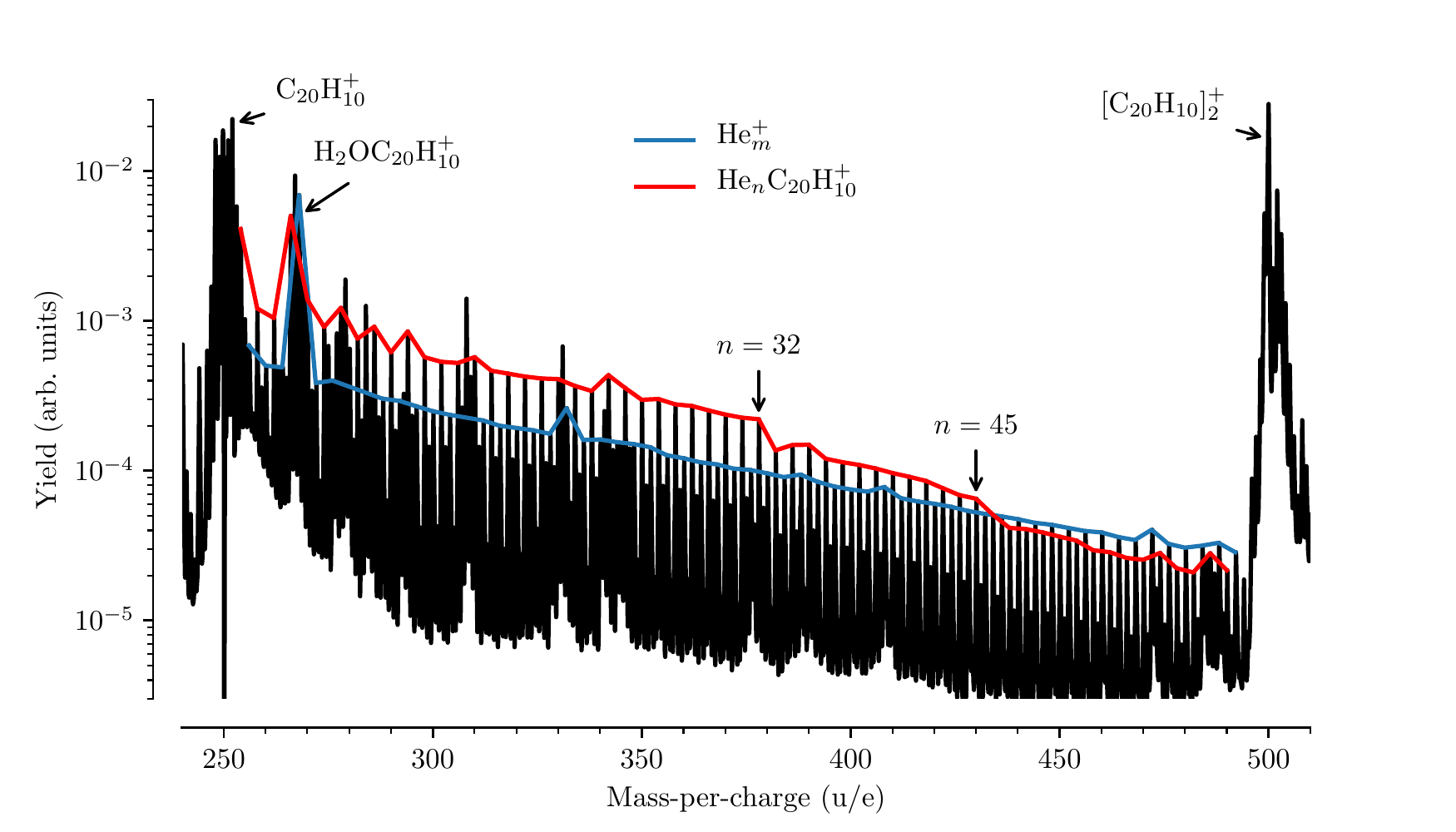} 
   \caption{Mass spectrum of corannulene cations formed by electron impact ionization of doped He nanodroplets. The series of He-complexed corannulene ions are indicated by the red line and a blue line shows the pure He$_m^+$ series. There are a number of contaminants that give additional features in the mass spectrum, the most prominent one being water giving a strong signal of H$_2$OC$_{20}$H$_{10}^+$ near mass 268\,u. Note the clear intensity drops in the red He$_n$C$_{20}$H$_{10}^+$ curve after $n=32$ and $n=45$.} 
   \label{fig:CorHeMS}
\end{figure*}

While detailed information about the solvation of corannulene in He can be deduced from the spectroscopy of He$_n$C$_{20}$H$_{10}^+$ complexes, these features are for the most part not visible in our mass spectra from these measurements. In Figure \ref{fig:CorHeMS} a mass spectrum obtained in these measurements to determine the depletion of He$_n$C$_{20}$H$_{10}^+$ complexes as function of laser wavelength is shown. The two strongest features in the mass spectrum are the peaks of the corannulene monomer and dimer. Between these we detect the series of pure He$_m^+$ clusters ($63 \leq m \leq 124$), indicated by the blue trace in Figure \ref{fig:CorHeMS}, and the He$_n$C$_{20}$H$_{10}^+$ series for $1 \leq n \leq 60$ indicated in red. In addition, a number of contaminants litter the mass spectrum with the most abundant one coming from the corannulene cation complexed with a single water molecule (or O or OH fragments) near 268 u/e. These contaminants are mainly introduced via the corannulene sample being used, but at least some of the water could also be picked up from the residual gas in our device. The pure He$_m^+$ cluster series follows an exponential decrease in intensity as a function of size (hence the linear appearance on the logarithmic scale), with a few deviations being the result of overlapping peaks from contaminants (the largest one being caused by the aforementioned H$_2$OC$_{20}$H$_{10}^+$ complex). The He$_n$C$_{20}$H$_{10}^+$ series follows a similar exponential decrease in intensity up to $n=32$ (with deviations again mainly being caused by contaminants), and again up to $n=45$, before clear drops in intensity are measured. These changes in intensity are typical of a shell closures, in this case the filling of the first solvation shell in two steps, and similar features have been found for He$_n$C$_{60}^+$ at $n=32$ and $n=60$.\cite{PhysRevLett.108.076101,Kuhn:2016aa} Interestingly, the second kink in the exponential intensity curve visible after $n=45$ is not observed in the absorption spectroscopy data (Figure \ref{fig:trend}).

Comparing the data in Figures \ref{fig:trend} and \ref{fig:CorHeMS} illustrates a practical side effect of performing action spectroscopy of ions solvated in He droplets: An increased sensitivity to the structures of loosely bound complexes over pure mass spectrometry and a decreased sensitivity to contaminants of similar mass as the system being studied. The sensitivity of the spectroscopic measurements to small, chemically induced shifts has allowed us to identify features that give a detailed insight into the solvation of corannulene ions in He, whereas the mass spectrum alone only reveals the complete closure of the first solvation shell. This method could thus be applied for experimentally studying the structures of small nanoparticles, clusters, and molecules. As noted in the introduction, action spectroscopy using He tags is also a sensitive tool for performing spectroscopic studies on complex molecular systems at temperatures close to 0\,K and with a much smaller perturbation than with traditional matrix isolation techniques. Also, since it is unlikely that two very different systems have electronic transitions at exactly the same energy, the contaminants that make the mass spectrum difficult to analyze will generally be transparent to the photons that are at resonance with a transition in the studied He$_n$C$_{20}$H$_{10}^+$ complexes. The main effect that the contaminants have on our measurements is that they increase the background noise level, reducing the signal-to-noise ratio in systems that overlap in mass with a contaminant. However, this is only an issue for very strong features such as the H$_2$OC$_{20}$H$_{10}^+$ complex which, as can be seen in Figure \ref{fig:CorHeMS}, is an order of magnitude more intense that the He$_n$C$_{20}$H$_{10}^+$ series at that mass. Additionally, since we measure the depletion of all masses as a function of laser wavelength at the same time with high mass resolution, we can be sure that we are indeed only measuring the absorption from the desired He-complexed ions.

\subsection*{Absorption spectra of protonated corannulene complexes}
By introducing D$_2$ into the pickup chamber of our experimental setup, we form He droplets doped with a mixture of deuterium and corannulene. Upon ionization, excess energy often leads to the breakup of at least one D$_2$ molecule and the formation of protonated C$_{20}$H$_{10}$D$^+$ systems. By using D$_2$ instead of H$_2$ we obtain a larger mass separation from the C$_{20}$H$_{10}^+$ radical, and avoid overlap with its $^{13}$C$^{12}$C$_{19}$H$_{10}^+$ isotopologue which makes up about one out of every six corannulene molecules. Like He, intact D$_2$ molecules can form weakly bound solvation layers around molecules such as corannulene. Previous studies with coronene and C$_{60}$ cations solvated in H$_2$/D$_2$ and He have shown that hydrogen molecules form closed solvation shells at the same numbers as helium.\cite{Leidlmair:2011aa,Goulart:2017aa} 

\begin{figure*}[t] 
   \centering
   \includegraphics[width=7in]{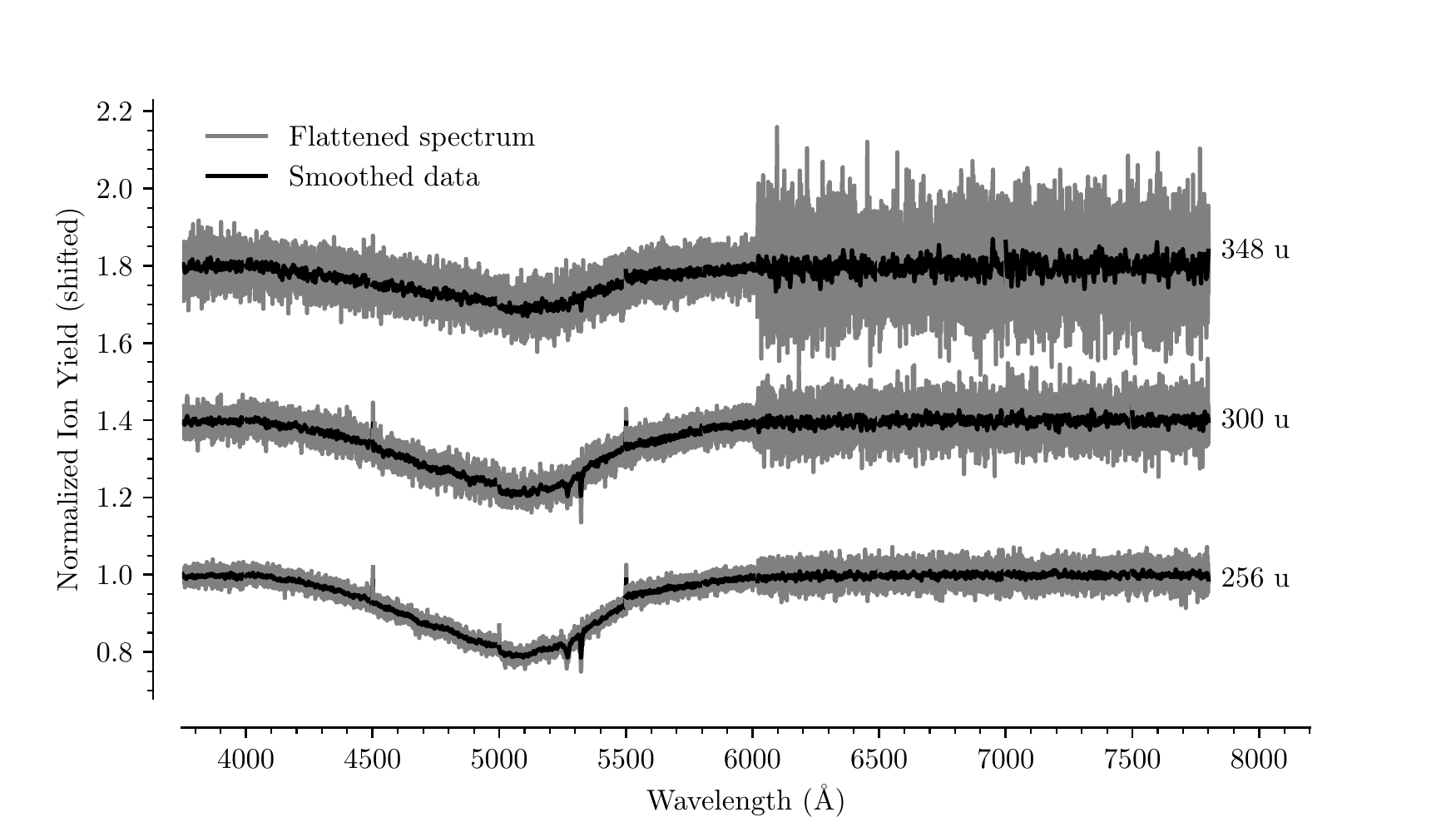} 
   \caption{Absorption spectrum in the range from 3750\,\AA\ to 7750\,\AA\ of C$_{20}$H$_{10}$D$^+$ complexed with 1, 12, and 24 He/D$_2$ atoms/molecules, respectively. The flattened raw spectrum is shown in gray and is overlaid with a spectrum where the data has been smoothed (moving average over 10 nearest neighbors) for easier viewing. A broad absorption feature is observed between 4000\,\AA\ to 6000\,\AA\ together with two narrow lines near 5280\,\AA\ and 5320\,\AA. The positions of the latter do not shift as a function of the solvation and are caused by an increased output from our laser that we have not been able to fully correct for. The measurements at wavelength above 6000\,\AA\ were performed with shorter dwell times and have a higher signal-to-noise ratio than at shorter wavelengths. A small discontinuity is also visible where two datasets were spliced together at 5500\,\AA.} 
   \label{fig:DOV}
\end{figure*}

Figure \ref{fig:DOV} shows the absorption spectrum of C$_{20}$H$_{10}$D$^+$ measured by the depletion of ions with masses of 256, 300, and 348\,u. Due to the similar masses of intact D$_2$ and He, we do not differentiate between the different mixtures of the two in the present measurement since both are weakly interacting and can be used as tags for action spectroscopy. The main feature in the absorption spectra is the very broad band between 4000\,\AA\ and 6000\,\AA. In addition, two narrow lines can be seen near near 5280\,\AA\ and 5320\,\AA, the positions of which do not change as a function of the number of solvating molecules/atoms. These are due to an enhanced output of our OPO laser at the wavelength emitted by the pump laser (5325\,\AA\ is the second harmonic of the pump laser), and thus an enhancement of the already present depletion at these wavelengths, that we not been able to fully correct for.

The protonated corannulene ion has a dramatically different absorption spectrum than the radical corannulene cation. Gone are the numerous narrow absorption bands and we only detect a single broad feature. Since this absorption feature is so broad, we are not able to identify any shift in its position as a function of the number of solvating He/D$_2$ units. The position of this band is in agreement with the S$_3$($^1$A) $\leftarrow$ S$_0$($^1$A) transition of protonated corannulene that was measured using a resonant multiphoton fragmentation technique in an ion trap as a broad peak near 5150\,\AA.\cite{Rice:2015aa} The width of this peak, FWHM $\approx800$\,\AA, can be explained by a short-lived S$_3$($^1$A) state with a lifetime of $\approx 2$\,fs, though unresolved vibrational excitations could contribute to the width making this lifetime a lower limit estimate. With no narrow bands originating from protonated corannulene in our measurements, it is unlikely that this system is responsible for any DIBs in this wavelength range. However, broad features like this could contribute to features in interstellar extinction curves like those that have been proposed to originate from hydrogenated fullerene molecules.\cite{Webster:1997aa}

\section*{Summary and Conclusions}

We have used helium nanodroplets to produce corannulene cations solvated by He that we have measured the absorption spectra of. This method is highly efficient in producing cold molecular samples with weakly interacting tags for action spectroscopy whereby the absorption spectrum of the tagged molecule is measured indirectly from the depletion of He-complexed systems as a function of the laser wavelength. From our measurements we identify 13 absorption bands between 5500\,\AA\ and 6000\,\AA\ and we have accurately studied the perturbation to the band positions as a function of the number of solvating He atoms. The behavior is found to be much more erratic than the linear behavior seen with He$_n$C$_{60}^+$ complexes up to $n=32$. The different features seen with corannulene ions solvated in He are a sensitive tool for studying the structures of the weakly bound complexes and the results agree very well with our computational modeling, revealing features not seen when using mass spectrometry alone. By introducing D$_2$ to our He droplets before ionization, we form protonated corannulene complexed with He and intact D$_2$. With the protonated system we do not detect any narrow absorption bands that could be potential DIBs candidates, but instead observe a broad absorption feature between 4000\,\AA\ and 6000\,\AA.

A disadvantage with the types of measurements performed here is that they are not directly accessing the gas phase band positions. This is a general problem with all forms of spectroscopy where a matrix or tags are used. However, He interacts only very weakly with other systems, minimizing these effects. So for systems with only one He atom attached, the chemical shift to the positions of absorption bands should be no more than a few {\AA}ngstr\"{o}ms. Given the relatively strong binding energy of the first He atom to the corannulene cation (about 27\,meV) compared to the C$_{60}$ cation (about 10\,meV\cite{PhysRevLett.108.076101}), the chemical shift induced to the electronic transitions induced by the first He atom that binds to corannulene is likely larger than the 0.7\,\AA\ seen with C$_{60}^+$,\cite{Kuhn:2016aa,Campbell:2016ab} perhaps greater than 1\,{\AA}. Although small for most applications, this shift is important to consider when comparing the experimental data with accurate astronomical measurements.

Finally, the ability to produce a wide range of cold mixed complexes in He nanodroplets and to perform action spectroscopy on them is a powerful tool for studying the structures of molecular ions and small nanoparticles. We foresee this being a useful method not only for investigating potential carriers of the diffuse interstellar bands, but for many other applications as well for systems where gas phase spectroscopy is difficult to perform, and as a highly sensitive tool for probing the structures of nano-solvated complexes.


\section*{Conflicts of interest}
There are no conflicts to declare.

\section*{Acknowledgments}
This work is supported by the Austrian Science Fund FWF (project P31149), and the Swedish Research Council (Contract No. 2016-06625). PM gratefully acknowledges financial support from the University of Innsbruck for this project.



\balance


\bibliography{Library.bib}
\bibliographystyle{rsc} 

\end{document}